\documentclass[a4paper, 12pt]{article}
\usepackage{graphicx,wrapfig,lipsum}
\usepackage{authblk}
\begin{document}

\title{Central Exclusive Production at RHIC}

\author[1]{Leszek Adamczyk}
\affil[1]{
Faculty of Physics and Applied Computer Science \\
AGH - University of Science and Technology \\
Krak\'ow, Poland\\
Leszek.Adamczyk@agh.edu.pl}

\author[2]{W\l odek Guryn}
\affil[2]{Brookhaven National Laboratory\\
Upton, NY 11973
USA\\
guryn@bnl.gov}

\author[3]{Jacek Turnau}

\affil[3]{Institute of Nuclear Physics\\
Krak\'ow, Poland\\
Jacek.Turnau@ifj.edu.pl}

\maketitle
\begin{abstract}
The present status and future plans of the physics program of Central Exclusive Production (CEP) at RHIC are described. The measurements are based on the detection of the forward protons from the Double Pomeron Exchange (DPE) process in the Roman Pot system and of the recoil system of charged particles from the DPE process measured in the STAR experiment's Time Projection Chamber (TPC). The data described here were taken using polarized proton-proton collisions at $\sqrt{s}=200~\textrm{GeV}$. The preliminary spectra of two pion and four pion invariant mass reconstructed by STAR TPC in central region of pseudo-rapidity $|\eta| < 1$, are presented. Near future plans to take data with the current system at center-of-mass energy $\sqrt{s}=200~\textrm{GeV}$ and plans to upgrade the forward proton tagging system are presented. Also a possible addition of the Roman Pots to the sPHENIX detector is discussed.
\end{abstract}


\section{Introduction}
Diffractive processes at high energies are believed to be occurring via the exchange of a color singlet object (the ``Pomeron'')  with internal quantum numbers of the vacuum~\cite{pomeron}. Even though properties of diffractive scattering at high energies are described by the phenomenology of Pomeron (${I\!\!P}$) exchange in the context of Regge theory, the exact nature of the Pomeron still remains elusive.  The main theoretical difficulties in applying QCD to diffraction are due to the intrinsically non-perturbative nature of the process in the kinematic and energy ranges of the data currently available.  In terms of QCD, Pomeron exchange consists of the exchange of a color singlet combination of gluons. Hence, triggering on forward protons at high (RHIC) energies selects exchanges mediated by the gluonic matter. 
\par
One of the main motivations for the inelastic diffraction program is searching for a gluonic bound state (glueball) whose existence is allowed in pure gauge QCD, but for which no unambiguous candidate has been established~\cite{glueball}. Idea that glueballs might be preferentially produced in the DPE process due to high gluon density in such process can be traced back to  ~\cite{Robson}. Two of the gluons in the DPE process could merge into a mesonic bound state without a constituent quark, forming a glueball in the central production process $pp \rightarrow pXp$. 
Indeed, the experimentally measured glueball candidates for the scalar glueball states are the $f_0(1500)$ and the $f_J(1710)$~\cite{abatziz}  in CEP as well as other gluon-rich reactions such as $\bar{p}p$ annihilation, and radiative $J/\psi$ decay~\cite{glueball_exp}. However, the energy regime where centrally produced glueball candidates have been identified so far is estimated to be not DPE dominated~\cite{klempt}. Thus it is imperative to cover a wide kinematic range to extract information of the production of glueball candidates at an energy regime where DPE is expected to be a dominant process.

The experiments at CERN ISR Collider \cite{AFS,ABCDHW1,ABCDHW2} and CERN SPS  \cite{WA91,WA102} have provided measurements of many CEP-type processes, however their interpretation in terms of Pomeron-Pomeron interactions is not fully justified \cite{LSpaper} at these rather low center-of-mass energy (62~\textrm{GeV} for ISR and 30~\textrm{GeV} for SPS).

\begin{figure}[htbp]
  \includegraphics[height=.17\textheight]{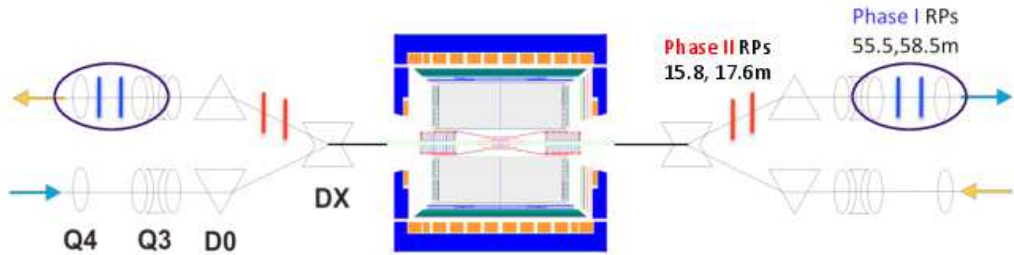}
  \caption{\small{The layout of the RPs with the STAR detector (not to scale). 
   The Phase I setup, designed to detect scattered protons with low-$t$, 
  is located after two dipole magnets (DX, D0) and three quadruples at 55.5 m and 58.5 m from the interaction point (IP), 
  respectively. For measuring protons with high-$t$ (Phase II), sets of RPs will be  positioned  between DX and D0 magnets, at 15.8 m and 17.6 m from IP.}  
  } \label{Fig:RPatSTAR}
\end{figure}

The Relativistic Heavy Ion Collider (RHIC) with its polarized proton beams provides a unique opportunity to  extend these studies to center-of-mass energies ($\sqrt{s}$) up to 510~\textrm{GeV} and investigate  spin dependence of diffractive scattering of polarized protons. There are two major experiments currently taking data at RHIC: STAR~\cite{star} and PHENIX~\cite{phenix}. At this time only the STAR experiment has the capability to explore physics related to the CEP. In $pp$ collisions, the outgoing forward protons are detected in the forward detectors, Roman Pots (RP), while the small number of charged tracks of the recoil system are measured in the STAR central detector, whose main part is the Time Projection Chamber (TPC).
Tagging and measuring the forward protons is important since it removes the ambiguity of a (complementary) rapidity gap tag. Furthermore, the momentum balance between the scattered protons and the centrally produced system, exclusivity condition, allows obtaining a relatively background free data sample, like the one obtained in year 2009  at $\sqrt{s} = 200~\textrm{GeV}$ (see section \ref{PhaseI}).
In the 2009 year set-up (Phase I) momenta of the forward protons are not reconstructed, therefore capability to apply energy-momentum conservation constraints on selected events is limited to very small momentum transfers. The program at STAR described here utilizes RP system of the pp2pp experiment~\cite{pp2ppNIM}, which was installed downstream of the STAR detector at RHIC, see Fig.~\ref{Fig:RPatSTAR}, where two locations of the Roman Pots are shown: current Phase I location and Phase II location being implemented to take data in 2015. The program will study both elastic and inelastic processes in a wide kinematic range in the energy range of RHIC $\sqrt{s}$ up to $510~\textrm{GeV}$ ~\cite{ProposalPhaseII}. 

\section{Current Status: CEP of pion pairs - Phase I setup at STAR}
\label{PhaseI}

The Central Production data were collected during five-day period of running using special optics of $\beta^*=20~\textrm{m}$.  The events were required to have two outgoing protons in the RPs, and the  tracks in the central region which were reconstructed with STAR Time Projection Chamber (TPC) covering $-1<\eta<1$ . The sub-sample with two opposite charge tracks matched with proper signals in Time Of Flight (TOF) detector has been fully analyzed and preliminary results reported recently \cite{DIS2014}. The exclusivity condition for such events is the transverse momentum balance  $p_T^{miss} <0.02 ~\textrm{GeV}$ between all four particles 
in the final state.\par
\begin{wrapfigure}{l}{5.5cm}
\includegraphics[width=5.5cm]{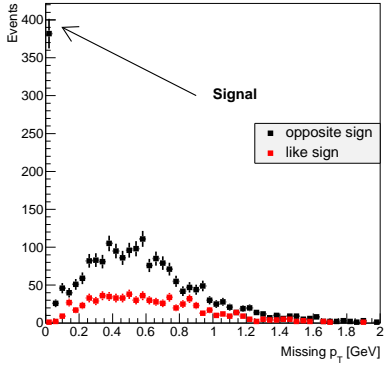}
\caption{\small{Transverse momentum balance distribution for events selected by RP trigger and requirement
of two tracks with TOF signal. Unlike (like) charge sign events are indicated by black (red) data points.}}\label{fig1:ptbalance}
\end{wrapfigure} 

The transverse momenta of final protons are estimated assuming no energy loss of scattered protons.
It was verified that in the range of the invariant $\pi^+\pi^-$ mass $M_{\pi\pi} < 1.5 ~\textrm{GeV}$ the effect of proton momentum loss is negligible for the   $p_T^{miss} < 0.02 ~\textrm{GeV}$ condition. In Fig. \ref{fig1:ptbalance} number of selected unlike sign (black points) and like sign (red points) meson pairs as a function of the $p_T^{miss}$ is shown. It can be seen that the non-exclusive background, the source of like sign events, quickly decreases with decreased transverse momentum imbalance and the above mentioned cut leaves practically clean sample of 380  CEP events.   

The visible kinematical range for which the acceptance and detector effciency corrections have been made is the following: proton four momentum transfer  squared $0.005 < -t_1, -t_2 < 0.03 ~\textrm{GeV}^2 $, pions pseudorapidity $\vert \eta_{\pi} \vert < 1$ and pseudorapidity of the $\pi\pi$ system $\vert  \eta_{\pi\pi}\vert < 2.0$.
STAR preliminary cross section for CEP of $\pi^+\pi^-$ at $\sqrt{s}=200 ~\textrm{GeV}$ in defined above visible kinematic range is 
$133 \pm 8(\rm{stat.)} \pm 12(\rm{syst.})  \rm{nb} $.

\begin{wrapfigure}{l}{5.5cm}
\includegraphics[width=5.5cm]{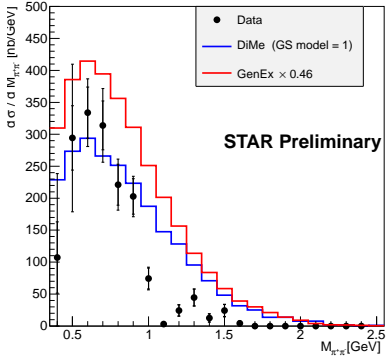}
\caption{\small{Differential cross section for exclusive production of $\pi^+\pi^-$ pairs as a function of their invariant mass $M_{\pi\pi}$. Error bars represent statistical and systematic uncertainties added in quadrature. Prediction of GenEx with off-shell form-factor parameter $\Lambda_{off}^2=1.6 ~\textrm{GeV}^2$ and survival factor calculated in \cite{Lebphd} $\textrm{SF}=0.46$ is shown as red curve. 
Prediction of DiMe Model 1 is represented as blue curve.}}
\label{fig4:invmass}
\end{wrapfigure}

\begin{figure}[bt]
\centering
\begin{minipage}[b]{0.4\linewidth}
\centering
  \includegraphics[width=\textwidth]{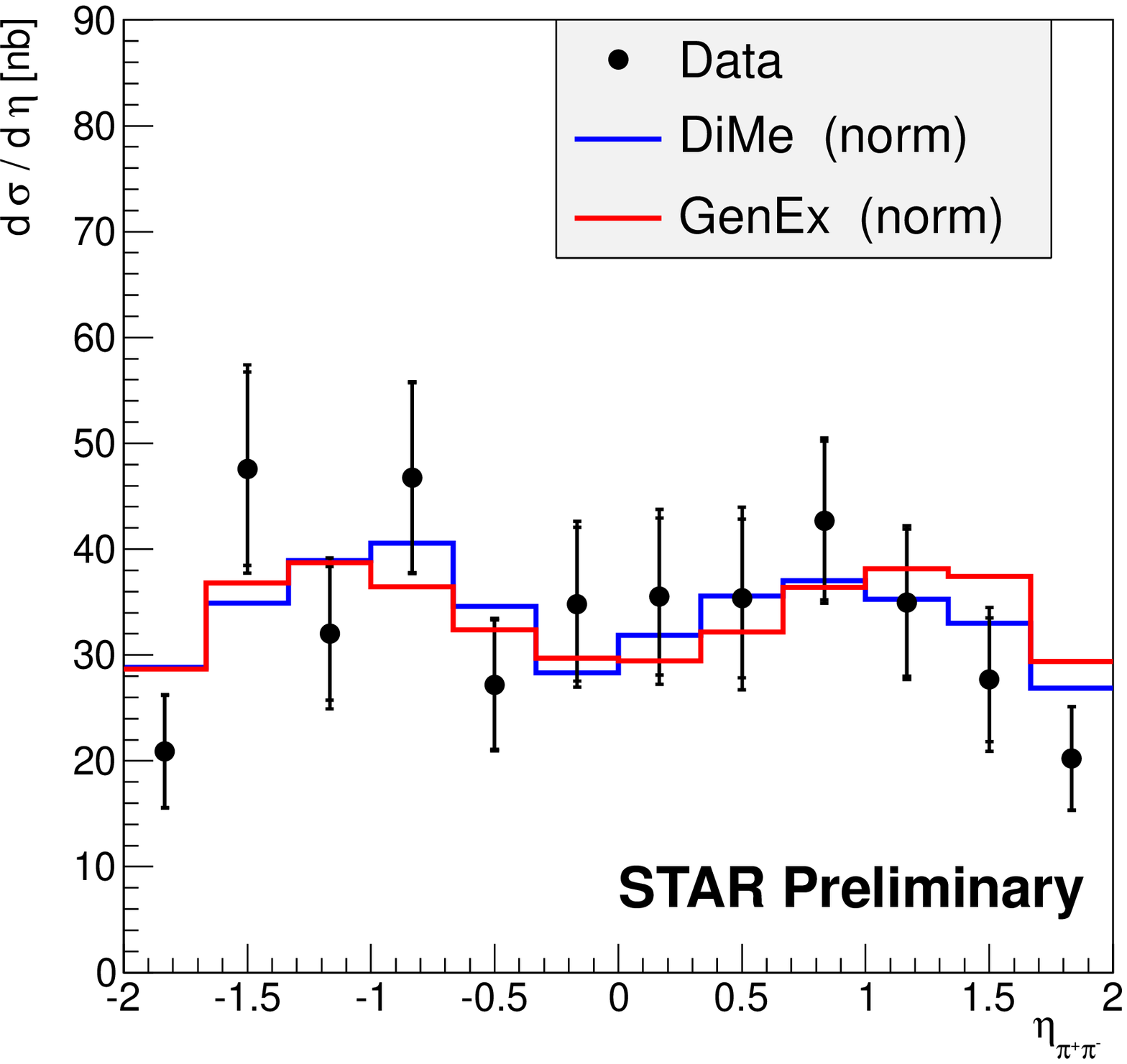}
\caption{{\footnotesize Differential cross section for the exclusive production of $\pi^+\pi^-$ system as a function of its pseudorapidity. Red and blue lines represent results of GenEx and DiMe respectively.$~~~~~~~~~~~~~~~~
~~~~~~~~~~~~~~~~~~~~~~~~~~~~~$}}
\label{fig4:pipieta}
\end{minipage}
\quad
\begin{minipage}[b]{0.4\linewidth}
\centering
  \includegraphics[width=\textwidth]{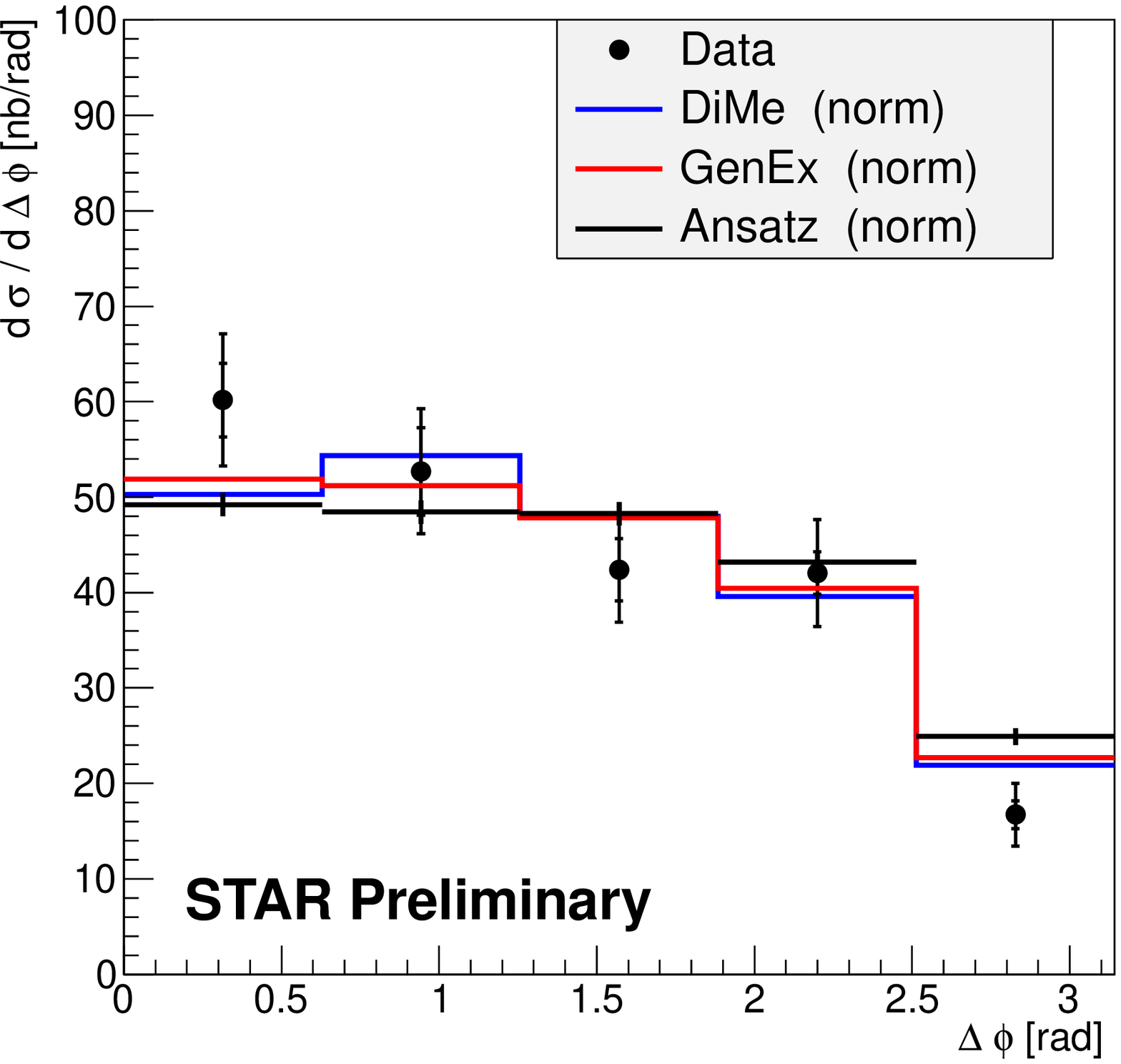}
\caption{{\footnotesize Differential cross section for the exclusive production of $\pi^+\pi^-$ system as a function of azimuthal angle between scattered protons. Red and blue and black lines represent results of GenEx, DiMe and Ansatz respectively}}
\label{fig5:deltaphi}
\end{minipage}
\end{figure}

In figure \ref{fig4:invmass} preliminary result on differential cross section for $pp \rightarrow p\pi^+\pi^-p$ process is presented as a function of $M_{\pi\pi}$. The $\pi\pi$ invariant mass spectrum shows characteristic features observed in CEP experiments at lower energies quoted in the Introduction: wide $f_0(600)$ resonance, sharp drop around 1 ~\textrm{GeV} due to interference of $f_0(980)$ with non-resonant
 background and some structure in the region 1.2 - 1.5 ~\textrm{GeV}. Present low statistics does not allow for any meaningful interpretation of the resonance structure however broad test consistency of the non-resonant background models \cite{LSpaper} and \cite{DIME} with measured differential cross sections is possible. By far the largest uncertainties in tested models are: the unknown off-shell form-factor of the Born term amplitude and soft survival factor (SF) due to screening corrections. 
The off-shell form-factor can be parametrized e.g. as $e^{(t-m_{\pi}^2/\Lambda_{off}^2)}$ where $\Lambda_{off}^2$ is the parameter which controls peripherality of the process and can be tuned to the data. In DiMe~\cite{DIME} and GenEx~\cite{GenEx} MC generators off-shell form-factor was adjusted in such a way that the non-resonant background in the mass spectrum presented in \cite{AFS} is nowhere above the data points. In GenEx  $\Lambda_{off}^2=1.6 ~\textrm{GeV}^2$ while for DiMe, in which the SF is much smaller,  $\Lambda_{off}^2=2.2 ~\textrm{GeV}^2$.  
The SF calculated for kinematic range of this experiment in ~\cite{Lebphd}  and Born term defined in ~\cite{LSpaper} is $~0.46$. DiMe prediction for $SF$ is $~0.26$. In figure ~\ref{fig4:invmass} predictions of GenEx with $SF=0.46$ and DiMe are shown. It can be seen that the parameter  $\Lambda_{off}^2$ affects strongly not only the normalization but also the shape of the invariant mass spectrum. 
Since the resonant contributions models do not describe the invariant mass distribution above 1 ~\textrm{GeV} all the other distributions, e.g. Fig.\ref{fig4:pipieta}, are calculated in the range $M_{\pi\pi} < 1 ~\textrm{GeV}$ and predictions are normalized to the cross section measured in this range. Both models describe the data well, however it should be noted that the agreement of similar quality has been obtained much simpler generator, called "Ansatz" in the legend of figure \ref{fig5:deltaphi}, which does not include any dynamical correlations between protons. We conclude that the shape of distributions of kinematical variables (except for $M_{\pi\pi}$), including azimuthal correlation between the scattered protons shown in figure \ref{fig5:deltaphi}, is completely determined by kinematical constraints and assumption of S-wave dominance in the decay of the centrally produced system. Both tested models are good candidates for non-resonant background in the exclusive production of meson pairs.

\section{Future Program: CEP - Phase II setup at STAR}

The next data taking for CEP and other diffractive topics is planned with the Phase II* RP setup in 2015 with polarized proton beams at $\sqrt{s} = 200 ~\textrm{GeV}$~\cite{STARBUR15}. Proton-nucleus (pA) collisions are also foreseen during that run including proton tagging. 

The expected CEP yields at $\sqrt(s) = 200 ~\textrm{GeV}$ are ~150K exclusive $\pi\pi$, with ~15K in $M_{\pi\pi} > 1 ~\textrm{GeV}$. The invariant mass range is up to 3 ~\textrm{GeV}. 
Those yields are based on assumed 200 Hz DAQ rate for CEP trigger.

Data taking in 2016 at RHIC (Run 16) at $\sqrt{s} = 510 ~\textrm{GeV}$ is also planned. This will allow a wider $t$-range (0.21 $< -t < 1.5 ~\textrm{GeV}^2$).

Addition of the Roman Pots to the long term upgrade of the PHENIX detector to sPHENIX is~\cite{sPHENIX} also planned. The sPHENIX detector plans to have a good jet measurement capability will allow study of jet production in CEP process.

\section{Conclusions and outlook}
We have described current and future diffractive physics program in polarized $pp$ collisions at RHIC with tagged forward protons using  the STAR detector at RHIC.  The program will study diffraction process in the RHIC $\sqrt s$ range up to 510 ~\textrm{GeV}. It will explore both elastic and inelastic  diffraction and search for predicted by QCD glueball. This diffractive program is complementary to the other RHIC physics program, will help understand both the strong interaction and  the hadronic structure within the framework of QCD.

The measurement of the central exclusive production of $\pi^+\pi^-$ pairs in proton-proton collisions at $\sqrt{s}=200 ~\textrm{GeV}$ at RHIC demonstrates exceptional efficiency of the Roman Pot tagging of scattered protons for reduction of the non-exclusive background. Cross section for this process measured in the visible kinematic range with $~15\%$ uncertainty is broadly consistent with predictions of DiMe ~\cite{DIME} and GenEx ~\cite{GenEx} generators, based on Regge phenomenology and tuned to the ISR collider CEP data at $\sqrt{s}= 62 ~\textrm{GeV}$. Except for the $M_{\pi\pi}$ invariant mass spectrum, shape of the measured distributions is well described by models, however it should be noted that effect of kinematic constraints is dominant. In particular substantial azimuthal angle correlation between scattered protons has purely kinematic origin.

The measurements of the Central Exclusive Production process will also allow study of gluon-gluon coupling through particle production. Heavy flavors could be produced. For example $\chi_C$ meson production cross section in the decay channel $\chi_c\rightarrow J/\psi + \gamma$ has been calculated~\cite{Durham2010} and is estimated to be 0.57 $nb$ at $\sqrt s= 500 ~\textrm{GeV}$ in the center of rapidity. Other groups~\cite{SzczurekPasechnik} estimated total production cross section, including absorption in NLO QCD corrections and gap survival probability to be about 5 $nb$ at $\sqrt s=200 ~\textrm{GeV}$. In the Pomeron Odderon interaction $J/\psi$ could also be produced as calculated in~\cite{Bzdak}.

\par
Preparations for CEP  measurements in STAR 2015 run are in progress. Present RP setup will be moved to new position in which RP proton tagging will be possible with the standard beam optics i.e. during normal STAR running. In effect we expect data with 30 - 40 times of present statistics in the higher and wider range of four-momentum transfers to scattered protons. The Phase II* vacuum chamber design will accommodate horizontal Roman Pots to increase the acceptance of the forward protons.
\section*{Acknowledgments}
This work was supported in part by the Office of NP within the U.S. DOE Office of Science and by the Polish National Science Centre under contract UMO-2011/01/M/ST2/04126.


\end{document}